\documentclass[12pt]{article}
\usepackage{arxiv}
\usepackage{url}
\usepackage{mathtools}
\mathtoolsset{showonlyrefs=true}
\usepackage{hyperref}
\usepackage{natbib}
\usepackage{graphicx}
\usepackage{amsmath}
\usepackage{xcolor}
\usepackage[english]{babel}
\usepackage{algorithm}
\usepackage[noend]{algpseudocode}
\usepackage{titlesec}
\usepackage{bbm}%
\usepackage{amsfonts}%
\usepackage{amssymb}
\usepackage{tikz}
\usetikzlibrary{positioning}
\usepackage{bm}
\usepackage{booktabs}
\usepackage{tikz}
\usepackage{enumitem}
\usepackage{amsmath}       
\usetikzlibrary{bayesnet}
\usepackage [autostyle, english = american]{csquotes}
\providecommand{\U}[1]{\protect\rule{.1in}{.1in}}
\newcommand{\beq}{\begin{equation}}
\newcommand{\eeq}{\end{equation}}
\newcommand{\bea}{\begin{eqnarray}}
\newcommand{\eea}{\end{eqnarray}}
\newcommand{\ba}{\begin{array}}
\newcommand{\ea}{\end{array}}
\newcommand{\bi}{\begin{itemize}}
\newcommand{\ei}{\end{itemize}}
\newcommand{\ben}{\begin{enumerate}}
\newcommand{\een}{\end{enumerate}}

\newcommand{\yy}{\mathbf{y}}
\newcommand{\xx}{\mathbf{x}}
\newcommand{\zz}{\mathbf{z}}

\newcommand{\ff}{\mathbf{f}}

\newtheorem{assumption}{Assumption}
\titleformat{\subparagraph}
  {\itshape}{\thesection}{1em}{}

\begin{document}

\title{Evaluating the Impact of COVID-19 Vaccination in the United Kingdom: A Gaussian Process Approach}
\author{Gianluca Giudice, Sara Geneletti and Konstantinos Kalogeropoulos\\
 Department of Statistics, London School of Economics\\
  \texttt{g.giudice@lse.ac.uk}; \texttt{k.kalogeropoulos@lse.ac.uk};\texttt{s.geneletti@lse.ac.uk};}
\maketitle

\begin{abstract}
The rapid rollout of COVID-19 vaccines in the United Kingdom in early 2021 differed markedly from that of many other European countries, providing a natural setting to assess the impact of vaccination speed on public health outcomes. We evaluate the impact of the accelerated UK vaccination rollout and associated policy transition on COVID-19 mortality and transmission dynamics by constructing a probabilistic reference trajectory for the UK under a slower vaccination and reopening trajectory. The proposed framework combines ideas from interrupted time series analysis and synthetic control methods with flexible probabilistic modelling based on multi-output Gaussian processes. These models capture non-linear and heterogeneous dependence structures across countries and over time, while providing uncertainty quantification through predictive distributions. A central feature of the methodology is a design-consistent validation strategy based on predictive performance in held-out pre-intervention periods, which is used both to guide model specification and to assess the plausibility of the reconstructed reference trajectory. The empirical results indicate a substantial reduction in COVID-19 mortality associated with the accelerated vaccination-policy transition, with little evidence of an effect on transmission rates. Generally, the framework illustrates how flexible probabilistic models and predictive validation can support causal and policy evaluation in complex time series settings.
\end{abstract}

\section{Introduction}

The rollout of COVID-19 vaccines varied substantially across countries, both in timing and in speed. In early 2021, the United Kingdom implemented one of the fastest vaccination programmes in Europe, reaching high levels of coverage earlier than many comparable countries \citep{Mathieu2021, Hale2021}. This divergence provides a natural setting to study whether differences in vaccination speed translated into measurable differences in public health outcomes. While a large body of clinical and epidemiological evidence shows that vaccination reduces the risk of severe illness and death \citep{Zheng22, Dagan2021, LopezBernal2021, Watson2022}, the extent to which differences in vaccination speed and associated policy responses affected outcomes at the population level, such as mortality and transmission dynamics, remains less clear.

Understanding these effects is not straightforward. Aggregate outcomes reflect a combination of factors, including behavioural responses, non-pharmaceutical interventions, and the evolution of the virus itself. Importantly, the accelerated vaccination rollout in the UK coincided with an earlier relaxation of restrictions and broader behavioural changes relative to many comparable European countries. As a result, the analysis should be interpreted as evaluating the broader vaccination-policy transition observed in the UK, rather than the isolated biological effect of vaccination alone. Moreover, interventions aimed at reducing severe outcomes may have more immediate and detectable effects on mortality than on transmission dynamics, which are often more sensitive to behavioural responses, population mixing, and the emergence of new variants \citep{Eyre2022, Bobrovitz2023}. Hence, differences in vaccination speed may translate more clearly into reductions in mortality than into changes in transmission rates. Beyond the specific COVID-19 setting, understanding these effects remains important for future epidemic preparedness and policy evaluation. A growing policy literature has emphasised the need to understand which interventions were most effective in order to inform preparedness for future epidemics \citep{WHO2024, WorldBank2025}, while recent empirical work continues to revisit vaccine effectiveness as new data become available \citep{Andrews2024}.

Addressing these questions in a time series setting presents additional challenges. In the absence of a direct comparison group, inference relies on constructing a credible reference trajectory for the treated unit, representing how outcomes would have evolved in the absence of the intervention. This requires combining information from other units and covariates while accounting for temporal dynamics and uncertainty. In practice, conclusions depend critically on modelling choices and on how well the model reproduces observed patterns prior to the intervention. Existing approaches often rely on relatively restrictive parametric assumptions governing temporal dynamics and cross-unit relationships. Misspecification of these structures may lead to poor reconstruction of the untreated trajectory even when the underlying causal assumptions are approximately satisfied.

A large literature has developed approaches for causal inference in time series and panel settings, including difference-in-differences methods \citep{Angrist2009, Callaway2021, Sun2021}, synthetic control approaches \citep{AbadieEtAl2010}, and Bayesian structural time series models \citep{Brod2014}. Recent work has also emphasised the importance of carefully defining intervention assumptions and validating untreated trajectory reconstruction in observational policy evaluation settings \citep{Menchetti2023, Papadogeorgou2023}. More flexible probabilistic approaches, including Gaussian process models, have also been proposed to capture complex relationships and heterogeneous dependence structures in observational data \citep{Alaa2017, BenMichael2023}. Related questions have emerged in epidemic modelling, where recent work has increasingly emphasised formal causal and policy evaluation perspectives within Bayesian infectious disease models \citep{Bhatt2023}, together with broader discussion regarding the role of statistical modelling and causal interpretation in epidemic policy evaluation \citep{Wood2026}.

In this paper, we evaluate the impact of the UK vaccination-policy transition on COVID-19 mortality and transmission dynamics using a Gaussian process framework. The proposed approach combines pre-intervention UK dynamics with information from comparable European countries in order to construct a probabilistic reference trajectory for the UK under a slower vaccination and reopening trajectory. Building on ideas from synthetic control and Bayesian time series modelling, the framework employs multi-output Gaussian processes to flexibly model non-linear and heterogeneous dependence structures across countries and over time, while providing coherent quantification of uncertainty through the corresponding predictive distributions. The framework is motivated by the observation that causal inference in these settings depends critically on the ability to reconstruct the untreated trajectory of the treated unit. Accordingly, flexible multi-output Gaussian Process models are combined with a validation strategy based on held-out pre-intervention data, which is used both to guide model specification and to assess the plausibility of the resulting reference trajectory. Applying the framework to weekly data on deaths and reproduction rates, we find evidence of a substantial reduction in mortality associated with the accelerated vaccination rollout, while effects on transmission are less clear.

Beyond the specific application, the paper highlights the importance of predictive validation in settings where causal conclusions depend on reconstructing the post-intervention untreated trajectory of a treated unit. More broadly, the paper illustrates how flexible probabilistic forecasting frameworks can support policy evaluation in complex observational time series settings while reducing reliance on restrictive parametric assumptions. The remainder of the paper proceeds as follows. Section 2 introduces the causal framework and estimands. Section 3 presents the Gaussian process models, while Section 4 describes the validation strategy and estimation procedure. Section 5 reports the empirical results, and Section 6 concludes.

\section{Causal Framework}
\label{sec:Background}

The application we refer to throughout the paper serves as an illustrative example and is analysed in Section \ref{sec:EmpiricalAnalysis}. It attempts to assess whether the accelerated vaccination rollout and associated earlier relaxation of restrictions in the UK affected mortality and transmission dynamics during the first semester of 2021. Formally, the treated unit is the UK and the intervention corresponds to the broader vaccination-policy transition associated with the substantially accelerated vaccination schedule. Other EU countries, with slower vaccination campaigns, are used to construct a reference trajectory for the UK. In this sense, we compare the observed trajectory of the UK with that of a non-treated version of itself under a slower vaccination rollout, constructed using other European countries.

We now introduce the notation and data structure used throughout the paper. Each observation is denoted by $y_{i,t}\in \mathcal{Y}$, where $i=1,\cdots, m$ indexes countries and $t = 1, \cdots, T_i$ indexes time. Each observation is associated with a set of $d$ potentially time-varying predictors ${\xx_{i,t}} \in \mathcal{X}^d$ such that
\begin{equation}
y_{i,t} = f(\xx_{i,t}) + \epsilon_{i,t},  \;\; \epsilon_{i,t}
\sim\mathcal{N}(0,\omega^{2}_i), \label{General_equation}
\end{equation} 
where $f(\cdot)$ is a generic function expressing the input-output relationship and $\epsilon_{i,t}$ is an error term with mean 0 and variance $\omega_i^2$. The $d$-dimensional feature vector ${\xx_{i,t}}$ is a set of time series specific to each unit $i$. In our application this includes mobility data and number of tests for each country. The data span $T_i$ periods and the first $t_0$ periods correspond to the data before the intervention, i.e. when the vaccination campaign began in the UK. 

\subsection{Assumptions}
\label{Assumptions}

In this subsection, we set up the framework to estimate the causal effect of an intervention on the treated unit. Each observation $y_{i,t}$ is associated with a binary potential outcome $y_{i,t}(w_{i,t})  \in \mathbb{R}$, where $w_{i,t} \in \{0,1\}$ is a treatment indicator, with 1 corresponding to treatment (the UK) and 0 to control units. Furthermore, define $\bm w_{1:m,1:T}$ as the assignment path up to time $T$ for all units.

Following recent work on causal inference and policy evaluation in time series settings \citep{Menchetti2021b, Menchetti2023, Papadogeorgou2023}, we formulate a set of assumptions under which differences between observed and reference trajectories may be interpreted as evidence of intervention effects. Since some assumptions are not directly testable, their credibility must ultimately be assessed in light of the empirical setting and the model's ability to reproduce pre-intervention dynamics.

\begin{assumption}[Single intervention]
Unit $i$ receives a single intervention at time $t_0$, such that $w_{i,t} = 0$ for $t < t_{0}$ and $w_{i,t} = w_i$ for all $t > t_{0}$.
\label{Assumption1}
\end{assumption}

This assumption reflects the nature of the vaccination programme, which is introduced at a specific point in time and persists thereafter without reversal. 

\bigskip

\begin{assumption}[Temporal no-interference]
For all $i$ and all $t > t_0$, the outcome of unit $i$ depends only on its own treatment path.
\end{assumption}

This is the time series analogue of the Stable Unit Treatment Value Assumption (SUTVA) \citep{Rubin1974}. In our empirical setting, vaccination policies are implemented at the country level and, during the period analysed, mobility restrictions were substantial. As a result, cross-country interference in outcomes such as deaths or infection rates is plausibly limited over the period considered. While some degree of cross-country spillover cannot be entirely ruled out—for example through travel or shared epidemiological dynamics—such effects are likely to be limited over the period considered. In particular, widespread mobility restrictions and largely country-specific policy responses reduce the scope for substantial interference across units.

\bigskip

\begin{assumption}[Covariates-treatment independence]
For $t > t_0$, covariates are not affected by the intervention.
\label{Assumption3}
\end{assumption}

This assumption ensures that covariates can be used for prediction without introducing post-treatment bias. In our application, variables such as testing intensity and mobility are treated as exogenous to the speed of the vaccination rollout over the period considered. In particular, while vaccination may indirectly influence behaviour, we assume that such effects are limited over the short horizon studied and do not materially affect the predictive role of these covariates. This assumption is primarily intended to avoid using variables whose post-intervention evolution may themselves be driven by the treatment.

\bigskip

\begin{assumption}[Non-anticipating outcomes]
For all $i$ and $t < t_0$, the outcomes are independent of the future intervention.
\end{assumption}

This assumption rules out anticipation effects. Although vaccination campaigns were announced in advance, there is no clear evidence that mortality or transmission dynamics systematically adjusted in anticipation of differences in rollout speed across countries.

\bigskip

\begin{assumption}[Stability of untreated dynamics]
The relationship between outcomes, covariates and latent temporal dynamics governing the untreated trajectory remains stable across the pre- and post-intervention periods.
\label{Assumption5}
\end{assumption}

This assumption implies that the untreated post-intervention trajectory can be inferred from pre-intervention dynamics together with the observed evolution of the control units. In the present setting, it means that, absent the accelerated vaccination rollout, the UK would have continued to evolve according to patterns comparable to those observed before the intervention and across the selected reference countries. Similar stability assumptions have recently been discussed in Bayesian policy evaluation settings by \cite{Papadogeorgou2023}. Since this assumption is not directly testable, the validation framework introduced in Section \ref{sec:Methodology} plays an important role in assessing whether the model can reproduce untreated dynamics during held-out pre-intervention periods.  

\subsection{Causal Estimands}
\label{subsection:Causal_Estimands}

Within our empirical application, these quantities summarise the effect of a faster vaccination rollout on mortality and transmission dynamics over time. Let $\delta_{i,t} =  y_{i,t}(1) - y_{i,t}(0)$ denote the causal effect at time $t$. The conditional average effect is
\begin{equation}
\tau_{i,t} = \mathbb{E}\left(\delta_{i,t}  | \xx_{i,t} \right).
\end{equation}

We are also interested in uncertainty quantification, which can be expressed via the variance or through credible intervals derived from the distribution of $\delta_{i,t}$. In addition to pointwise effects, we consider the cumulative effect of the intervention over time
\begin{equation}
\mathcal{T}_i = \sum_{t=t_0+1}^{T_i}\tau_{i,t},
\end{equation}
which is particularly meaningful for flow variables such as deaths. For stock variables, such as the reproduction rate, it is more appropriate to consider the average effect
\begin{equation}
\overline{\tau}_i = \frac{1}{T_i-t_0}\sum_{t=t_0+1}^{T_i}\tau_{i,t}.
\end{equation}

Within a GP framework, these quantities can be obtained from the posterior predictive distribution. When appropriate, transformations (e.g. logarithmic) can be used to express effects on a multiplicative scale and then mapped back to the original scale for interpretation.

\section{Gaussian Process Models and Benchmark Specifications}
\label{sec:Gaussian Process}

Most existing approaches to policy evaluation in time series settings rely on parametric specifications for the relationship between outcomes, covariates and temporal dynamics. In this paper, we adopt a Gaussian Process (GP) framework, which allows these relationships to be modelled non-parametrically while providing a coherent probabilistic description of uncertainty \citep{Rasmussen06}. In our setting, the role of the model is to construct a reference trajectory for the treated unit under no intervention. Rather than imposing a specific functional form, the unknown relationship between outcomes and predictors is represented through a prior distribution over functions defined by a covariance kernel. Inference and prediction are then based on the corresponding posterior and predictive distributions. We consider three classes of models that differ in how they exploit information from control units and represent temporal dependence. The primary framework is a multi-output Gaussian Process, while a single-output Gaussian Process and a Bayesian Structural Time Series model are included as benchmark specifications.

\subsection{Overview and modelling strategy}

The objective of the analysis is to construct a credible reference trajectory for the treated unit, representing how outcomes would have evolved under a slower vaccination and reopening trajectory. In this setting, causal conclusions depend critically on the quality of the reconstructed untreated trajectory. We therefore consider several modelling approaches that differ in how they exploit information from control units and represent temporal dependence.

The primary framework considered in this paper is a \emph{multi-output Gaussian Process} (MOGP), which models the trajectories of all countries jointly. By explicitly representing dependence across countries, MOGPs allow information from control units to contribute directly to the reconstruction of the reference trajectory for the treated unit. This framework also accommodates non-linear relationships between outcomes and covariates and provides a coherent probabilistic description of uncertainty. To assess the value of modelling cross-country dependence explicitly, we also consider a \emph{single-output Gaussian Process} (SOGP). In this specification, the outcome of the treated unit is modelled directly, while outcomes from control countries enter as predictors. This retains the flexibility of Gaussian Process regression but does not model the joint dependence structure across countries. Finally, we compare both GP-based approaches with the \emph{Bayesian Structural Time Series} (BSTS) framework of \cite{Brod2014}, which forms the basis of the Bayesian Causal Impact methodology. BSTS models represent a widely used benchmark for policy evaluation in time series settings and provide a useful comparison with more traditional state-space approaches.

The purpose of this comparison is not to identify a universally preferred model class, but rather to evaluate alternative specifications within a common validation framework. As discussed in Section \ref{sec:Estimation}, model selection is ultimately based on predictive performance in held-out pre-intervention periods. This allows the choice of model and control units to be guided by their ability to reconstruct observed untreated dynamics, rather than by a priori modelling preferences.

\subsection{Multi-output Gaussian Process}
\label{subsec:MultiO}

Multi-output GPs exploit correlations across outputs and inputs, which can improve predictive performance, particularly in the presence of noise or missing data \citep{Bonilla08}. In our setting, the model induces dependence across countries through shared latent processes, allowing information from control units to contribute directly to the reconstruction of the reference trajectory for the treated unit. 

Let $\yy = \{\yy_1^{'}, \cdots, \yy_m^{'}\}^{'}$, where $\yy_i^{'} =  \{ y_{i,1}, \cdots, y_{i,T_i}\}$, and let $X = \{X_1^{'}, \cdots, X_m^{'}\}^{'}$, with $ X_i \in \mathbb{R}^{T_i\times d}$ the covariates associated with unit $i$. We allow for settings in which different units may be observed at different time points and with different covariates \citep{Liu2018}. The model is given by
\begin{equation}
 y_{i,t} =  f_i(\xx_{i,t}) +  \epsilon_{i,t}, \qquad \epsilon_{i,t}
\sim\mathcal{N}(0,\omega^{2}_i), \label{GP}
\end{equation}
for $i = 1, \cdots , m$ and $t = 1, \cdots , T_i$. Stacking observations, the likelihood becomes
\begin{equation}
\yy |\ff (X),X,\Omega \sim \mathcal{N}(\ff(X),\Omega),
\end{equation}
with $\Omega = \text{diag}(\omega^2_{1} \text{I}_{T_1}, \cdots, \omega^2_{m} \text{I}_{T_m})$ and $\ff(X) = \{f_1(X_1), \cdots f_m(X_m)\}^{'}$.

We assume
\begin{equation}
\ff(X) \sim \mathcal{GP} (\bm 0 , \mathcal{K}(X, X)),
\end{equation}
where $\mathcal{K}(X, X)$ is a positive semi-definite covariance matrix encoding both temporal and cross-unit dependence. To specify $\mathcal{K}$, we adopt the Linear Model of Coregionalization (LMC), a widely used framework for multi-output Gaussian Processes \citep{Alvarez2012}. In the LMC, each output is expressed as a linear combination of latent Gaussian processes. The specification used here corresponds to a restricted low-rank version of the LMC, often referred to as the Semi-Parametric Latent Factor Model (SLFM) \citep{Teh2005}, in which
each output is expressed as a linear combination of shared latent processes:
\begin{equation}
f_i(X_i) = \sum_{q=1}^{Q} \lambda_{i,q} u_q(X_{i}),
\end{equation}
where $u_q(\cdot)$ are independent latent GPs and $\lambda_{i,q}$ are loadings capturing cross-unit dependence. This representation allows different countries to share latent temporal and covariate-driven components while retaining country-specific dependence through the loadings $\lambda_{i,q}$. The resulting covariance structure is
\begin{equation}
\mathcal{K}(X,X) = \sum_{q=1}^{Q} \text{B}_q \otimes K_q(X,X), \label{SigmaGP}
\end{equation}
where $\text{B}_q$ are coregionalisation matrices, and $K_q$ are covariance kernels defined on the input space. We also consider a specification combining a  covariate-driven kernel with a temporal kernel:
\begin{equation}
	\mathcal{K} =  \text{B}_1 \otimes K_{rbf}(X, X) + 
		  \text{B}_2 \otimes K_{Mat} (t,t), \label{SigmaF}
\end{equation}
where the first kernel operates on the observed covariates and the second on time. 

Unlike many epidemic models, which describe transmission through compartments and epidemiological parameters, the proposed framework is deliberately agnostic regarding the underlying transmission process. Instead, it focuses on reconstructing a reference trajectory for observed outcomes using information from covariates and comparable countries. The goal is therefore not to model disease dynamics directly, but to support causal evaluation through accurate prediction of untreated trajectories.

\subsection{Alternative benchmark specifications}

To assess the value of explicitly modelling cross-country dependence through a multi-output Gaussian Process, we also consider two alternative specifications that are commonly used for prediction and policy evaluation in time series settings. These models serve as benchmarks within the validation framework described in Section \ref{sec:Methodology}.

\subsubsection{Single-output Gaussian Process}

The first benchmark is a standard single-output Gaussian Process (SOGP), which models only the treated unit. Let
$
\yy = {y_1,\ldots,y_T}\in\mathbb{R}^T
$
denote the outcome series of the treated unit and let
$
X={\xx_1',\ldots,\xx_T'}'\in\mathbb{R}^{T\times d}
$
be the associated covariates. Furthermore, let
$
Z={\zz_1',\ldots,\zz_T'}'\in\mathbb{R}^{T\times(m-1)}
$
collect the outcomes of the remaining units. Defining the combined input vector
$
\xx_t^{*}=(\xx_t',\zz_t')',
$ the model takes the form
\begin{equation}
y_t=f(\xx_t^{*})+\epsilon_t,
\qquad
\epsilon_t\sim N(0,\omega^2),
\label{SOGP}
\end{equation}
where ($f(\cdot)$) is assigned a Gaussian Process prior with mean function $\mu(\cdot)$ and covariance function $k(\cdot,\cdot)$.

Unlike the multi-output framework, dependence across countries is not modelled explicitly. Instead, outcomes from the control units enter the model as predictors for the treated unit. This specification retains the flexibility of Gaussian Process regression while providing a useful benchmark for assessing the benefits of joint modelling across countries.

\subsubsection{Bayesian Structural Time Series}

Our second benchmark is the Bayesian Structural Time Series (BSTS) framework of \cite{Brod2014}, which forms the basis of the Bayesian Causal Impact methodology. Using the same data structure as the SOGP, the model is specified as
\begin{equation}
y_t=\mu_t+\bm{\beta}'\zz_t+\epsilon_t,
\qquad
\epsilon_t\sim N(0,\sigma^2_\epsilon),
\end{equation}
where $\zz_t$ contains the outcomes of the control units and $\mu_t$ is a latent trend component. Following \cite{Brod2014}, we employ a local linear trend specification,
\begin{align}
\mu_t &= \mu_{t-1}+\delta_{t-1}+\eta_t,\\
\delta_t &= \delta_{t-1}+\zeta_t,
\end{align}
with
\[
\eta_t\sim N(0,\sigma^2_\eta),
\qquad
\zeta_t\sim N(0,\sigma^2_\zeta).
\]

The BSTS model represents temporal dynamics through a parametric state-space formulation and has been widely used for intervention analysis and policy evaluation. It therefore provides a natural benchmark against which to compare the Gaussian Process approaches considered in this paper.

\section{Model specification, validation and inference}
\label{sec:Estimation}

\subsection{Overview}

We now describe the procedure used to specify, validate and estimate the models introduced in Section \ref{sec:Gaussian Process}. Our main concern is to ensure that the final model provides a credible reconstruction of the untreated trajectory of the treated unit before it is used for causal inference. To this end, all model specification decisions are based exclusively on pre-intervention data, so that the validation exercise is not contaminated by post-intervention outcomes. This ensures that model selection remains aligned with the objective of reconstructing the untreated trajectory rather than reproducing the observed treated outcomes.

The framework combines flexible predictive modelling with a validation strategy designed to assess whether a candidate model can reproduce untreated dynamics in periods where the outcomes are observed. This emphasis on predictive validation is motivated by Assumption \ref{Assumption5}, which requires stability of the untreated dynamics across the pre- and post-intervention periods. While no validation exercise can establish this assumption formally, poor predictive performance in held-out pre-intervention data would cast doubt on the ability of a model to reconstruct the untreated trajectory after the intervention. The use of flexible non-parametric models reduces reliance on restrictive assumptions regarding the functional form of the untreated trajectory, but also increases the importance of validation. Consequently, the same predictive distributions that are used to assess model performance in the pre-intervention period are subsequently used to construct the untreated reference trajectory and the associated causal estimands after the intervention.

There are two main issues to address. The first is the selection of appropriate control units. The second is the choice of model specification, including the dependence structure and kernel components. We treat both as part of a unified specification problem and evaluate candidate models through a placebo intervention exercise based on held-out pre-intervention data.

\subsection{Design-consistent validation framework}
\label{sec:Methodology}

The central challenge in prediction-based causal inference is that the quantity of interest, namely the untreated trajectory of the treated unit after the intervention, is inherently unobservable. Consequently, the credibility of the resulting causal conclusions depends critically on the ability of the model to reconstruct untreated dynamics. This is particularly important in flexible non-parametric settings, where predictive performance depends not only on the plausibility of the causal assumptions but also on the ability of the model to capture complex temporal and cross-unit relationships.

To assess the ability of each model to reconstruct the relevant dynamics, we introduce a pseudo-intervention time $t^* < t_0$ within the pre-intervention period and split the data into a training sample $t = 1,\ldots,t^*-1$ and a validation sample $t=t^*,\ldots,t_0$. In the empirical analysis, these correspond approximately to two thirds and one third of the pre-intervention observations, respectively. The logic is simple. Since the true intervention has not yet occurred, the outcomes observed after $t^*$ are known realizations under no intervention. We can therefore estimate each candidate model on the training sample, generate predictions for the validation period, and compare those predictions with the observed data. Models that are unable to reproduce untreated outcomes in the held-out pre-intervention period are unlikely to provide credible reconstructions of the untreated trajectory after the intervention.

This placebo intervention exercise also provides an empirical diagnostic for Assumption \ref{Assumption5}, which requires stability of the untreated dynamics across the pre- and post-intervention periods. While no validation exercise can establish this assumption formally, poor predictive performance would constitute evidence against the ability of the model to capture the underlying untreated dynamics and would therefore weaken confidence in subsequent causal conclusions. In this sense, the validation exercise serves not only as a model selection device but also as a diagnostic regarding the plausibility of the assumptions required for causal interpretation.

Restricting model specification and validation to the pre-intervention period is essential. Using post-intervention outcomes to select control units, tune model specifications, or compare competing models would amount to comparing observed treated outcomes $y_{i,t}(1)$ with predictions intended to approximate untreated outcomes $y_{i,t}(0)$. Such a procedure would favour models that produce trajectories closer to the treated observations and could mechanically shrink the estimated treatment effect towards zero. To avoid this source of bias, all model specification decisions are based exclusively on pre-intervention information. This perspective is closely related to recent Bayesian approaches to policy evaluation in time series settings, which emphasise the importance of untreated trajectory reconstruction and pre-intervention predictive performance as key ingredients for credible causal interpretation \citep{Papadogeorgou2023}. In the proposed framework, predictive validation plays a central role because the same predictive distributions used to assess model performance before the intervention are subsequently used to construct the untreated reference trajectory and the associated causal estimands after the intervention.

\subsection{Candidate models and control units}

The validation exercise requires comparing multiple model specifications and control-unit combinations. To keep this exercise computationally feasible, we first reduce the set of candidate control units using Dynamic Time Warping (DTW) \citep{Giorgino07}. DTW provides a measure of similarity between time series that remains informative when similar features occur at different times. This is particularly useful in epidemic applications, where countries may experience similar epidemic waves on different calendar schedules. Applying DTW to the pre-intervention period, we retain the eight countries with the smallest distances to the UK as the initial candidate control pool. Final control-unit selection is determined by predictive performance in the validation exercise.

Having identified a candidate pool of control units, we next consider the choice of model structure. We compare the following specifications:
\begin{itemize}
\item[1)] \textit{Two Factor GP} (2FGP). This is the model in \eqref{SigmaF}, where an rbf kernel is used on the covariate space and a Mat\'ern kernel on time. This corresponds to a two-factor Semi-Parametric Latent Factor Model, with one latent component capturing dependence on observed covariates and a second capturing residual temporal dependence. This formulation admits a latent factor interpretation and allows covariate-driven and temporal variation to be modelled separately
\item[2)] \textit{Independent GPs} (INGP). In this case the outputs are modelled independently by setting $\text{B}_q = \text{I}_m$, for $q=1,\ldots,Q$, so that no dependence is induced across countries. This specification provides a benchmark for assessing the value of borrowing information across units through the multi-output GP framework.
\item[3)] \textit{One Factor GP} (1FGP). Here we combine the rbf and Mat\'ern kernels on a common input space, so that $\mathcal{K} =  \text{B}_1 \otimes (K_{rbf}(X,X) +  K_{Mat}(t,t))$. This yields a simpler specification in which temporal and covariate effects are represented through a single shared latent component.
\item[4)] \textit{Two RBF Factor GP} (2RBF). As in the 2FGP, the input space is divided into temporal and covariate components, but both are modelled using rbf kernels: $\mathcal{K}=  \text{B}_1 \otimes K_{rbf}(X,X) + \text{B}_2 \otimes K_{rbf}(t,t)$.
\item[5)] \textit{Single Output Gaussian Process} (SOGP). This is the model in \eqref{SOGP}, where the inputs consist of the selected control-unit outcomes together with the covariates of the treated unit.
\item[6)] \textit{Bayesian Causal Impact} (BCI). This is the local linear trend model of \cite{Brod2014}, estimated using the same univariate data structure as the SOGP.
\end{itemize}

Models 1)–4) correspond to alternative specifications within the multi-output GP framework described in Section \ref{subsec:MultiO}. Model 5 provides a single-output GP benchmark, while Model 6 corresponds to the Bayesian structural time series approach of \cite{Brod2014}.

\subsection{Estimation}

Given a selected specification, estimation proceeds in two stages. During the model comparison stage, the number of candidate specifications is large, making a full Bayesian treatment computationally impractical. We therefore estimate hyperparameters using Type II maximum likelihood, obtained by maximizing the marginal likelihood
\begin{equation}
p(\yy|X, \bm \theta) = \int p(\yy| \ff,X,\bm \theta) p( \ff|X,\bm \theta),d\ff,
\end{equation}
where $\bm \theta = {\bm \phi^{'}, \bm \lambda^{'}, \bm \omega^{2'}}^{'}$ contains the kernel parameters, the coregionalisation parameters and the observation variances. Because both the likelihood and the GP prior are Gaussian, the integral is tractable and yields
\begin{equation}
\log p(\yy|X, \bm \theta) =
-\frac{1}{2}\yy'\Sigma^{-1}\yy
-\frac{1}{2}\log|\Sigma|
-\frac{T}{2}\log(2\pi),
\label{MLE}
\end{equation}
where $\Sigma=\mathcal{K}+\Omega$. In practice, optimisation is carried out using L-BFGS-B. Type II maximum likelihood provides a computationally efficient approach for comparing a large number of candidate specifications while retaining the flexibility of the GP framework. Once a single specification has been selected, we perform a Bayesian analysis in order to quantify parameter uncertainty. In the multi-output GP models, the parameters of primary interest are the loadings $\bm \lambda$ entering the coregionalisation matrices, since they govern the strength of dependence across countries and therefore play a central role in the reconstruction of the reference trajectory. Let $\bm \phi^*$ and $\bm \omega^{2*}$ denote the Type II maximum likelihood estimates of the kernel and variance parameters. We then approximate
\begin{equation}
p(\bm \lambda |X, \yy)=
\int p(\bm \lambda, \bm \phi, \bm \omega^2 | X, \yy)
, d\bm \phi , d\bm \omega^2
\approx
p(\bm \lambda | X, \yy, \bm \phi^*, \bm \omega^{2*}),
\end{equation}
thereby concentrating the computational effort on the parameters governing cross-country dependence. As a robustness check, we also consider freeing up $\bm \omega^2$ in order to account for uncertainty arising from the observation variances.

Posterior inference is performed using Hamiltonian Monte Carlo with weakly informative priors. In particular,
\begin{equation}
\lambda_i \sim \mathcal{N}(0,10^2),
\end{equation}
while, when estimated, the observation variances are assigned Gamma priors,
\begin{equation}
\omega_i^2 \sim \mathcal{G}(0.1,1).
\end{equation}

This stage allows uncertainty in the cross-country dependence structure to be propagated to the reconstructed reference trajectory and the resulting causal estimands.

\subsection{Predictive distributions, scoring rules and causal estimands}
\label{sec:prediction}

Predictive distributions play a dual role in the proposed framework. Before the intervention, they are used to assess model performance through predictive scoring rules. After the intervention, they define the reconstructed untreated trajectory and the associated causal estimands. This connection is central to the validation-centred approach developed in this paper, since the same predictive machinery is used both for model selection and for causal inference.

Under a Gaussian Process prior, the predictive distribution is available in closed form,
\begin{equation}
\widetilde{\yy} \mid \yy, X, \widetilde{X}
\sim
\mathcal{N}(\widetilde{\bm\mu},\widetilde{\Sigma}),
\end{equation}
where the predictive mean and covariance are given by the standard Gaussian Process expressions. When parameter uncertainty is incorporated, the predictive distribution is obtained by averaging over posterior draws of the relevant hyperparameters. During the validation stage, these predictive distributions are evaluated using scoring rules that assess different aspects of predictive performance. We consider the Mean Squared Error (MSE), the Logarithmic Score (LogS) \citep{Good52}, and the Energy Score (ES) \citep{Matheson76}. The MSE evaluates point prediction accuracy, while the LogS and ES assess both predictive accuracy and uncertainty quantification. For all three measures, lower values indicate better predictive performance. Candidate models and control-unit combinations are compared using their scores on the held-out pre-intervention validation sample, and the specification with the strongest overall predictive performance is selected for the causal analysis.

After the intervention time $t_0$, the predictive distribution is used to construct the untreated reference trajectory for the treated unit.

Under Assumptions \ref{Assumption1}--\ref{Assumption5}, the predictive distribution provides a probabilistic reconstruction of the untreated reference trajectory. Using predictive samples $\widetilde{y}_{i,t}^{[k]}(0)$, $k=1,\ldots,N$, we obtain samples from the posterior distributions of the pointwise, cumulative and average causal effects defined in Section \ref{subsection:Causal_Estimands}. Since the predictive distribution is joint over the post-intervention period, temporal dependence is naturally preserved when constructing these causal estimands.

\subsection{Practical workflow}

The proposed procedure can be summarised as follows.

\begin{itemize}
\item[$1)$] Remove countries with insufficient input or output data over the period of interest.

\item[$2)$] Restrict the analysis to the pre-intervention period and introduce a placebo intervention time $t^* < t_0$, thereby defining training and validation samples.

\item[$3)$] Apply DTW to the pre-intervention period in order to screen the available control units and retain a smaller candidate pool.

\item[$4)$] Fit all candidate specifications and select the one with the strongest predictive performance on the validation sample.

\item[$5)$] Re-estimate the selected specification on the full dataset, perform Bayesian inference, and obtain posterior predictive samples and the associated causal estimands.
\end{itemize}

Steps 1)--4) constitute the design-consistent validation stage and rely exclusively on pre-intervention information. Step 5 uses the selected specification to reconstruct the post-intervention reference trajectory and perform causal inference.

\section{Empirical analysis}
\label{sec:EmpiricalAnalysis}

\subsection{Application and data}

We study whether the accelerated vaccination rollout, and associated earlier reopening, implemented in the United Kingdom (UK) in early 2021 was associated with changes in COVID-19 mortality and transmission, relative to a reference trajectory constructed using European countries with slower vaccination rollouts. Evidence suggests that vaccination reduces the risk of severe outcomes, including death, and may also affect transmission dynamics, although the magnitude of this effect remains uncertain (see \cite{Zheng22} for a meta-analysis).

Our data consist of weekly observations for multiple European countries from $1^{\text{st}}$ March 2020 to $30^{\text{th}}$ June 2021. The intervention date $t_0$ is set to $31^{\text{st}}$ January 2021, corresponding to the first week in which the number of individuals receiving a second dose in the UK exceeded 500,000, marking the transition to large-scale full vaccination. The UK is treated as the exposed unit, while other European countries provide information for constructing a reference trajectory. We consider two outcome variables. The first is weekly confirmed COVID-19 deaths per million people, which we analyse on the logarithmic scale to mitigate skewness. The second is the effective reproduction number $R$, measuring the average number of secondary infections generated by an infected individual.

To improve predictive performance while remaining consistent with the covariates–treatment independence assumption (Assumption \ref{Assumption3}), we include the following covariates: (i) a time index, (ii) Google mobility data, and (iii) the weekly number of COVID-19 tests per 1,000 people. Mobility data are summarised using principal component analysis (PCA), retaining the first component, which explains on average over 80\% of the variation. The data are observed over different time spans across countries, with variation in start dates and covariate availability. Time is therefore indexed relative to the earliest available observation across countries, and the Gaussian process framework accommodates these differences through its covariance structure.

\subsection{Validation and model selection}

Following the validation framework of Section \ref{sec:Methodology}, all model specification decisions are based exclusively on pre-intervention data. First, we apply dynamic time warping (DTW) \citep{Giorgino07} to identify countries with similar pre-intervention trajectories to the UK. This reduces the pool of candidate control units to a subset of eight countries for each outcome. This step is intended to reduce the set of candidate controls to a manageable subset prior to the formal validation exercise, rather than to determine the final specification directly. Second, we split the pre-intervention period at a time $t^* < t_0$ into training and validation sets. Candidate specifications are compared using the predictive scores introduced in Section \ref{sec:prediction}.

\begin{table}[t]
\setlength{\tabcolsep}{15.3pt}
\centering
\begin{tabular}{rllllllllll}
\hline
\\[-12pt]
\multicolumn{1}{l}{} & \multicolumn{5}{c}{Weekly deaths per million people}  \\ 
\cmidrule(lr){2-7}
\multicolumn{1}{l}{} & 2FGP  & 1FGP  & 2RBF  & INGP  & SOGP   & BCI \\ \midrule
MSE  &  0.8189 &  0.8274 &  1.2754 &  4.2811 &  \textbf{0.7263} &  3.1693 \\
logS &  \textbf{0.2389} &  0.4624 &  0.3451 &  1.1465 &  0.4008 &  4.7445\\
ES   & \textbf{0.6776} &  0.7791 &  0.8820 &  2.7745 &  0.7025 &  2.6261  \\   \bottomrule \\[-12pt]
\multicolumn{1}{l}{} & \multicolumn{5}{c}{Weekly infection rate R}  \\ 
\cmidrule(lr){2-7}
\multicolumn{1}{l}{} & 2FGP  & 1FGP  & 2RBF  & INGP  & SOGP   & BCI \\ \midrule
MSE  & \textbf{0.3597} &  0.4679 &  0.4028 &  2.1470 &   0.4905 &  0.5386 \\
logS &  -0.6960 & -0.6815 & \textbf{-0.8182} &  0.7835 &   -0.6173 & -0.5928\\
ES   & \textbf{0.2883} &  0.3354 &  0.2915 &  1.4496 &   0.3505 &  0.3794  \\   \bottomrule
\end{tabular}
\caption{\small Comparison of predictive performance across candidate models. Lower values indicate better performance.}
\label{TableMSE}
\end{table}

\begin{figure}[t]
\centering
\includegraphics[scale = 0.45]{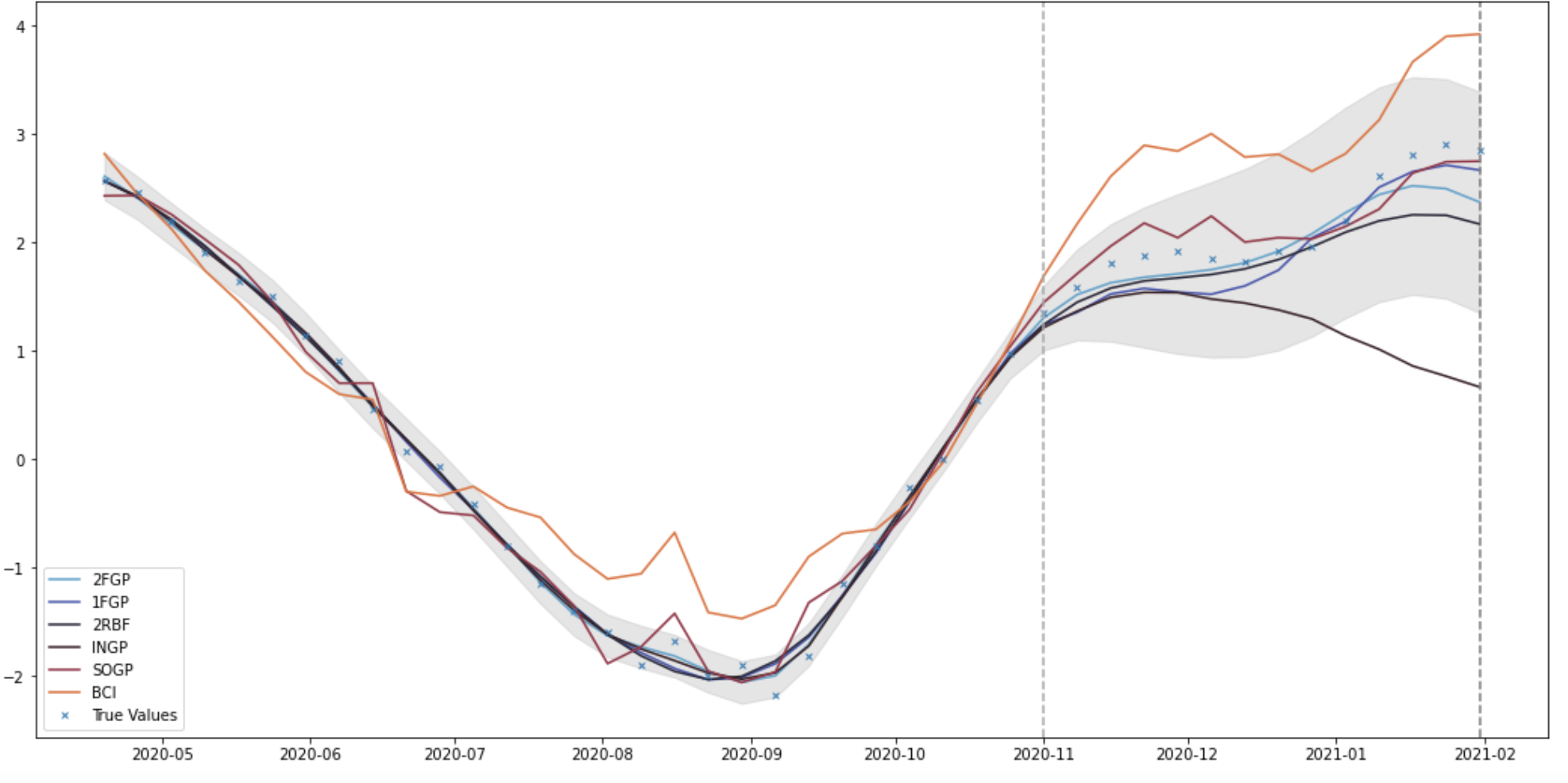}
\includegraphics[scale = 0.45]{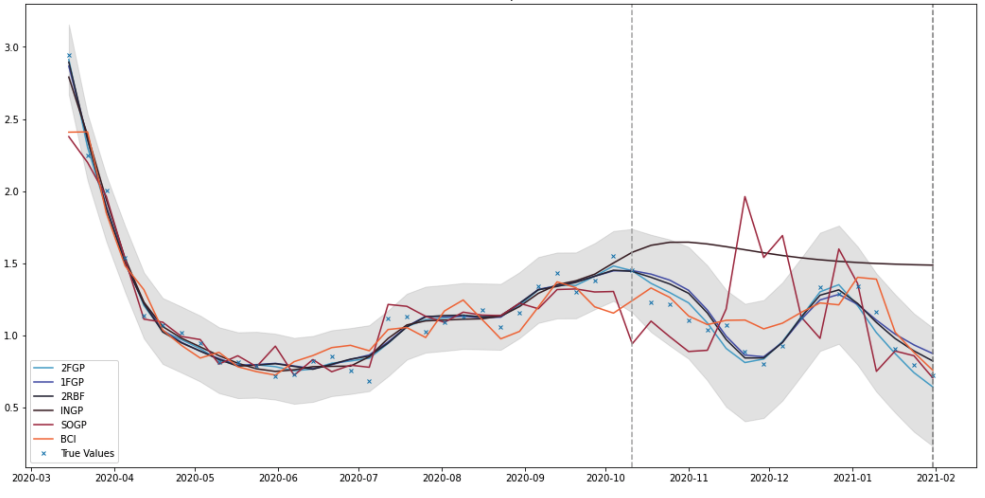}
\caption{\small Model comparison in the validation period. Top: log weekly deaths. Bottom: reproduction rate.}
\label{fig:ModelComparison}
\end{figure}

Table \ref{TableMSE} reports predictive performance across candidate specifications. The two-factor Gaussian process (2FGP) achieves the lowest energy score across both outcomes, indicating strong predictive performance in the pre-intervention period. Other GP-based models also perform well, while independent specifications (INGP) and the state-space benchmark (BCI) generally exhibit weaker predictive performance. Unlike synthetic control methods, which construct weighted combinations of control units, the Gaussian process framework models the joint distribution of outcomes across units, allowing for flexible non-linear and heterogeneous dependence structures together with coherent uncertainty quantification. Figure \ref{fig:ModelComparison} provides a visual comparison of model predictions over the validation period. Based on these validation results, the 2FGP specification is selected for the subsequent policy evaluation analysis.

\subsection{Estimated effects on mortality}

\begin{figure}[t]
\centering
\includegraphics[scale = 0.45]{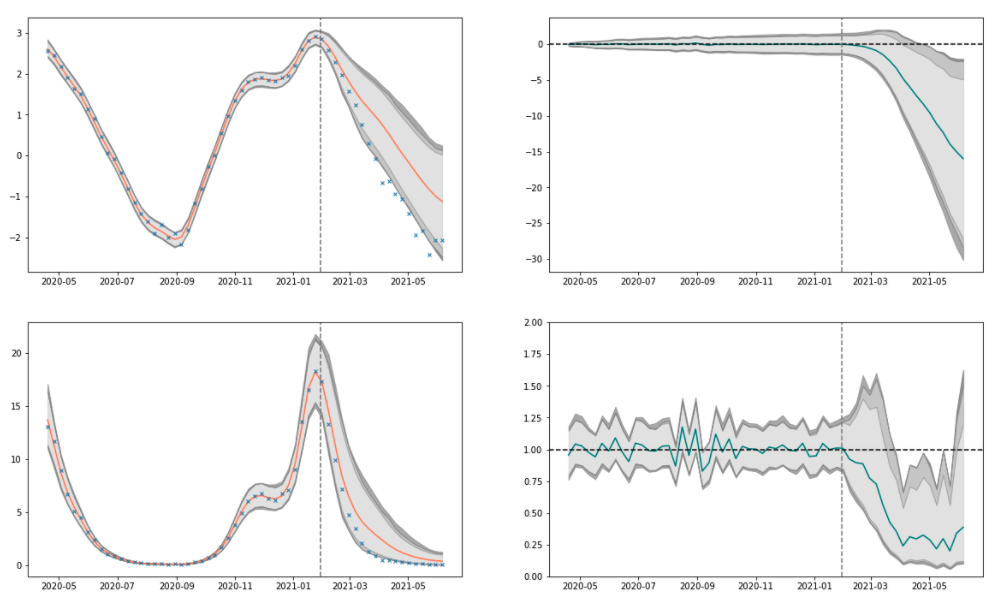}
\caption{\small Observed and predicted mortality trajectories and causal effects for the UK.}
\label{fig:DeathsUK}
\end{figure}

We re-estimate the selected model on the full dataset, restricting UK observations to the pre-intervention period. The selected control group consists of Italy, the Netherlands, Ireland, and Portugal. Figure \ref{fig:DeathsUK} shows observed and predicted trajectories. The model provides a close fit in the pre-intervention period, indicating that it captures the relevant dynamics of the data. The close alignment between observed and predicted trajectories prior to $t_0$ provides indirect empirical support for the plausibility of the constructed reference trajectory. The Gaussian process framework also allows us to account for parameter uncertainty. Incorporating uncertainty in the coregionalisation structure increases the width of the credible intervals but does not alter the qualitative conclusions.

After the intervention, the observed trajectory diverges from the predicted reference trajectory, with realised deaths consistently lower than predicted. The cumulative effect becomes increasingly negative after $t_0$, indicating lower mortality relative to the reference trajectory. The pointwise multiplicative effects suggest a substantial reduction. Averaging over the post-intervention period, the estimated multiplicative effect implies that mortality in the UK is approximately 51.4\% [30.1\%, 82.9\%] of the level predicted under the reference trajectory. This corresponds to a substantial reduction in mortality associated with the accelerated vaccination rollout and broader policy transition observed in the UK.

\subsection{Estimated effects on infection rates}

We repeat the analysis for the reproduction number $R$, using Portugal, Ireland, France, and Denmark as controls.
\begin{figure}[t]
\centering
\includegraphics[scale = 0.45]{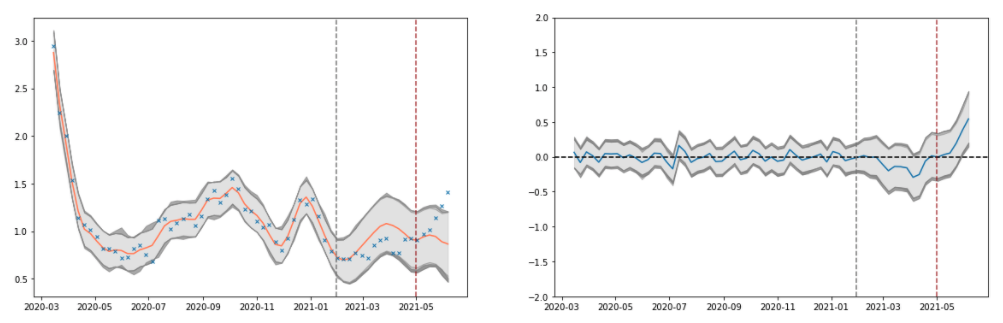}
\caption{\small Observed and predicted reproduction rate and causal effects.}
\label{fig:RoUK}
\end{figure}
As shown in Figure \ref{fig:RoUK}, the model provides a good fit in the pre-intervention period. After the intervention, deviations between observed and predicted trajectories are smaller and more uncertain than in the mortality case. The estimated average additive effect is $\overline{\tau}_i = 0.0063$ with 95\% credible interval $[-0.1582, 0.1653]$, indicating no clear evidence of a reduction in transmission. The results suggest that any effect on transmission is either small relative to variability in the data or difficult to detect given the observational setting and evolving epidemiological conditions. The divergence observed in late spring 2021 coincides with the period during which the Delta variant became dominant in the UK, which may partly explain the increasing discrepancy between the observed and reconstructed trajectories.

\subsection{Discussion}

The empirical analysis suggests that the accelerated vaccination campaign in the UK is associated with a substantial reduction in COVID-19 mortality relative to a reference trajectory constructed from European countries with slower rollouts. In contrast, we do not find clear evidence of a corresponding reduction in transmission. The credibility of these findings rests on the ability of the model to recover observed dynamics in the pre-intervention period. While this does not constitute a formal identification result, strong predictive performance in held-out pre-intervention periods provides an important diagnostic supporting the plausibility of the reconstructed untreated trajectory.

From a policy perspective, these results suggest that the timing and speed of vaccination rollout may play an important role in reducing severe outcomes, even when effects on transmission are more difficult to detect. As with most observational policy evaluations conducted during the pandemic, the estimated effects may also reflect broader contemporaneous epidemiological and policy developments that cannot be fully separated from the vaccination rollout. This reflects a broader challenge in epidemic policy evaluation, where interventions, behavioural responses, and epidemiological dynamics often evolve simultaneously and are difficult to disentangle completely. The analysis should therefore be interpreted as evaluating the impact of the accelerated vaccination strategy within the wider policy and epidemiological environment prevailing during the study period. At the same time, the findings should primarily be interpreted as comparative estimates relative to the selected reference group rather than as fully isolated causal effects, particularly given the potential for residual cross-country dependencies and shared epidemiological dynamics.

\section{Conclusion}

There is growing interest in evaluating the impact of interventions and policies in time series settings, particularly in the context of the COVID-19 pandemic. In this paper, we develop a Gaussian process-based framework for causal inference and policy evaluation in such settings. The starting point is Assumption \ref{Assumption5}, which requires that the untreated trajectory of the treated unit can be reconstructed from information contained in control units and observed covariates. This naturally shifts attention towards predictive modelling and raises the question of how the credibility of the reconstructed reference trajectory should be assessed.

Our contribution lies less in the introduction of a new predictive model and more in the integration of flexible multi-output Gaussian Processes within a principled validation-centred workflow for causal inference. Multi-output Gaussian Processes provide a flexible framework for modelling complex temporal and cross-unit dependence structures, while the proposed design-consistent validation strategy uses held-out pre-intervention periods to guide model specification and assess the plausibility of the resulting reference trajectory. While such validation does not establish identification in a formal causal sense, it provides an important design-based diagnostic linking the underlying causal assumptions to the predictive performance of the selected model.

We apply the framework to assess the impact of the accelerated COVID-19 vaccination rollout and associated policy transition in the United Kingdom. The results suggest a substantial reduction in mortality relative to a reference trajectory constructed from European countries with slower vaccination rollouts. In contrast, we do not find clear evidence of a corresponding reduction in the reproduction rate over the period considered. From a policy perspective, the results highlight the potential importance of the timing and speed of vaccination rollout in reducing severe outcomes during an epidemic. 

More generally, the multi-output Gaussian Process formulation admits a latent factor interpretation through its shared Gaussian Process components, suggesting connections with factor-based and matrix completion approaches to causal inference in panel and time series settings \citep{Athey2021}. Future work could extend the framework to settings with multiple interventions, allow explicitly for interference across units, incorporate more structured latent dependence formulations, or explore alternative validation strategies tailored to more complex empirical designs.

\bibliography{Ref}
\bibliographystyle{apalike}
\end{document}